\documentclass[reqno]{amsart}
\usepackage{graphicx}
\usepackage{graphics}
\usepackage{amsmath}
\usepackage{epsfig,bm}
\usepackage{rotating}
\usepackage{url}
\usepackage{float} 
\usepackage{amsthm}
	\newtheorem*{theorem}{Theorem}

\newcommand{\tsum}{\mbox{$\sum$}}

\DeclareMathOperator{\subs}{subs}
\DeclareMathOperator{\floor}{floor}

\begin{document}

\title[Monomer-Dimer Cluster Expansion]{
	Proof of Convergence for the Lattice Monomer-Dimer Cluster Expansion I, 
	a Simplified Model}

\author[P. Federbush]{P. Federbush\\
Department of Mathematics\\
	University of Michigan \\
	Ann Arbor, MI 48109-1043, USA}

\begin{abstract}
\renewcommand{\theequation}{\Alph{equation}}
We present some promising ideas to treat the problem of making completely rigorous the development 
of our expression for $\lambda_d(p)$ of the monomer-dimer problem on a $d$-dimensional 
hypercubic lattice
\begin{equation}\label{abstract1}
	\lambda_d(p)=\frac{1}{2}\Big(p\ln(2d)-p\ln(p)-2(1-p)\ln(1-p)-p\Big)
	+\sum_{k=2}a_k(d)p^k
\end{equation}
where $a_k(d)$ is a sum of powers $(1/d)^r$ for
\begin{equation}\label{abstract2}
	k-1\leq r\leq k/2
\end{equation}
In fact as we will point out one has allready rigorously established 
the convergence of the sum in \ref{abstract1} for small $p$. It is
the $d$ dependence of $a_k(d)$ that has yet to be rigorously shown. 
We do not now know how to complete the proof.
\end{abstract}

\maketitle

\renewcommand{\theequation}{\arabic{equation}}
\addtocounter{equation}{-2}
\section{The Model Problem}\label{s1}
We start by presenting the model problem. It is interesting in its own right,
and is after all the study of this paper.  $r>0$ is given and real $J_i$ are given,
$i\geq 2$ satisfying
\begin{equation}\label{1}
	|J_i|\leq r^i
\end{equation}
We define
\begin{equation}\label{2}
	Z=Z(N,p)=\sum_{\alpha_i}\prod_i\left((J_ip^iN)^{\alpha_i}\cdot\frac{1}{
	\alpha_i!}\right)
\end{equation}
The $\alpha_i,i\geq 2$ are non-negative integers, and 
the sum over the $a_i$ in \eqref{2} is over all values of the $\alpha_i$,
restricted by 
\begin{equation}\label{3}
	\sum_{i=2}^\infty i\alpha_i\leq \frac{pN}{2}
\end{equation}
We will prove

\begin{theorem}
	There is a $p_0>0$ such that for $0\leq p\leq p_0$
	\begin{equation}\label{4}
		\lim_{N\to\infty}\frac{\ln Z}{N}=\sum_{2}^\infty p^iJ_i
	\end{equation}
\end{theorem}

The easiest way to prove this theorem (probably) must
be to estimate the corresponding sum to that in \eqref{2} with the
$\alpha_i$ satisfying the complimentary inequality
\begin{equation}\label{5}
	\sum_{i=2}^\infty i\alpha_i>\frac{pN}2
\end{equation}
If this part is ``small enough'' then $Z$ becomes
approximately equal
\begin{equation}\label{6}
	Z\cong\prod_i\left(\sum_{\alpha_i}(J_ip^iN)^{\alpha_i}\cdot
	\frac{1}{\alpha_i!}\right)
\end{equation}
and the theorem is easily proved

We follow a much more devious route to the proof, in this paper. 
A route whose steps can all be paralleled in the actual problem.

\section{Introduction}\label{s2}
The cluster expansion approach to $\lambda_d$ of the dimer problem 
on a hypercubic lattice was presented in \cite{1}. A formal argument was 
given for the expansion
\begin{equation}\label{7}
	\lambda_d\sim\frac{1}{2}\ln(2d)-\frac{1}{2}+\sum_iC_i\frac{1}{d^i}
\end{equation}
At present the status of this putative asymptotic 
expansion is still
not clear. In \cite{4}, working with Friedland, a simple extension of the
expansion formalism of \cite{1} was made to treat the monomer-dimer problem. This 
yielded the expresion
\begin{equation}\label{8}
	\lambda_d(p)=\frac{1}{2}\Big(p\ln(2d)-p\ln p-2(1-p)\ln(1-p)-p\Big)+
	\sum_{k=2}a_k(d)p^k
\end{equation}
where $a_k(d)$ is a sum of powers of $(1/d),(1/d)^r,$ with 
\begin{equation}\label{9}
	k-1\geq r\geq k/2.
\end{equation} 

The situation is a little complicated. The cluster expansion formalism yields
an expression for $a_k(d)$ as a function of the Mayer Series
coefficients of the dimer gas. In \cite{4}, repeated in eq~(12)--(20) 
of \cite{3}, another route was obtained to derive the 
$a_k(d)$ from the mayer series coefficients. Although the two routes certainly 
give the same answer, this has not been rigorously established. The
inscrutable [WORDS] identity that must be proved in detail in \cite{5}. 
It is child's play to see that the second route leads rigorously
to an expression \eqref{8} for $\lambda_d(p)$ where the sum converges.
But it is the expression in \eqref{8} derived by the first route (still not
proved rigorously) for which the $a_k(d)$ have the expression in 
powers of $1/d$ described before \eqref{9}. Proof of the convergence of the 
cluster expansion, which we enter into in this paper, 
will show that the two expressions for all $a_k(d)$ are equal, that
the sum in \eqref{8} converges for small $p$, and that the 
$a_k(d)$ have the indicated dependence on $d$.

The limit that must be evaluated to rigorously establish the 
cluster expansion development is presented in Section~\ref{s3}. The relation of the
limit of the model problem, eq~\eqref{4}, to this limit will be clear. In
succeeding sections the theorem of eq~\eqref{4} is proven, taking care 
continuously to carry out steps as they can easily be applied to the general limit 
of Section~\ref{s3}. The basic strategy is to arrange $\ln Z$ as the sum
of terms, chunks. Within the chunks some sums are replaced by 
contour integrals (in many complex variables). A single stationary point
of the integrand [WORDS] in the limit of those integrals.

\section{The Object of Study}\label{s3}
We must analyze $Z^*$
\begin{equation}\label{10}
	Z^*=\sum_{\alpha_i}\beta(N,\mbox{$\sum$} i\alpha_i)\prod\bar{J}_i^{\alpha_i}
	\frac{N^{\sum\alpha_i}}{\prod(a_i!)}
\end{equation}
This is eq~(5.24) of \cite{2}. Here the $\alpha_i, i\leq2,$ are non-negative 
integers and are restricted by eq~\eqref{3}
\begin{equation}\label{11}
	\beta(N,jN)=e^{NH(p,j)}
\end{equation}
with 
\begin{equation}\label{12}
	\sum i\alpha_i=jN
\end{equation}
and
\begin{align}\label{13}
	H(p,j) & = j\ln p+(1-2j)\ln(1-2j)+j-\frac{p}{2}\left(1-\frac{2j}p\right)\cdot
		\ln\left(1-\frac{2j}p\right) \\ \label{14}
	& \equiv j\ln p+\tilde{H}(p,j)
\end{align}
Eq~\eqref{11} is eq~(5.16) of \cite{2}, eq~\eqref{13} is eq~(5.17) of 
\cite{2}. Eq~\eqref{14} defines $\tilde{H}$. We also define
$\tilde{\beta}$
\begin{equation}\label{15}
	\beta(N,\mbox{$\sum$}i\alpha_i)\equiv p^{\sum i\alpha_i}\tilde{\beta}(N,
	\mbox{$\sum$}i\alpha_i)\equiv p^{\sum i\alpha_i}e^{N\tilde{H}(p,j)}
\end{equation}
$Z^*$ becomes
\begin{equation}\label{16}
	Z^*=\sum_{\alpha_i}\tilde{\beta}(N,\tsum i\alpha_i)\prod (\bar{J}_ip^iN)^{\alpha_i}
	\frac{1}{\prod(\alpha_i!)}
\end{equation}
Sums are again restricted by \eqref{3}.

One wants to study
\begin{equation}\label{17}
	\lim_{N\to\infty}\frac{\ln Z^*}{N}
\end{equation}
The similarity between eq~\eqref{16} and eq~\eqref{17} and the
pair of equations, eq~\eqref{2} and eq~\eqref{4}, is obvious. The 
$\bar{J}_i$ have a weak dependence on $N$. (They are asymptotically constant.)

The desired limit of \eqref{17} we do not detail now. This limit might be found in
(5.31) and (5.32) of \cite{2}, or in an entirely different form in the discussion
surrounding (24)--(28) in \cite{3}, where another reference is given.

\section{From Sums to Contour Integrals}\label{s4}
In the next section $Z$ or $Z^*$ from \eqref{2} or
\eqref{16} will be arranged into a sum of terms called chunks. In
some of these chunks there will be a designated set of indices, 
$\mathcal{S}$ such that a portion of the chunk is of the form
\begin{equation}\label{18}
	\prod_{i\in\mathcal{S}}\left(\sum_{i=0}^{m_i}\frac{(J_ip^iN)^{\alpha_i}}
	{\alpha_i!}\right)\beta
\end{equation}
Here if it is $Z$ we are working with $\beta=1$, it is $Z^*$, $J_i$ becomes
$\bar{J}_i$; from now on such trivial differences will not be 
commented on. In \eqref{18} the other $\alpha_i$ are not summed, having
been assigned certain values and 
$\beta$ may depend on $\alpha_i$. The $J_i$ for $i$ in $\mathcal{S}$ will
be negative, it will be important to control cancellations between positive
and negative terms in evaluating \eqref{18} accurately enough. This motivated
the use of contour integrals. In fact dealing with $Z$ a simpler treatment 
is possible as will be pointed out later, but we want to use
a method that applies to both $Z$ and $Z^*$.

We set $-a\equiv J_iNp^i$ and note 
\begin{equation}\label{19}
	\sum_{\alpha=0}^n\frac{(-a)^\alpha}{\alpha!}f(\alpha)=
	\frac{1}{2\pi i}\oint_Cdz\frac{\pi}{\sin\pi z}\frac{a^z}{z!}f(z)
\end{equation}
where the contour $C$ is counterclockwise and contains $\{0,1,\dots,n\}$ 
and no other singularities of the integrand. Employing the identity
\eqref{19} in \eqref{18} for all the $\alpha_i$ with $i$ in $\mathcal{S}$
we have converted all the sums in our chunk to a single multivariable 
contour integral. Analytic properties of $\Gamma(z)$ and $\beta$ will be dealt
with later.

The division of $Z$ and $Z^*$ into chunks to convert sums from having limits as given by 
\eqref{3} to limits as in \eqref{19}
\begin{equation}\label{20}
	0\leq\alpha_i\leq m_i
\end{equation}
In the space of allowed $\alpha_i$ we are fitting hyper rectangles. This messy procedure
is not illucidated. It is we feel the central idea of the proof.

\section{Disection into Chunks}\label{s5}
We write this section in the language of $Z^*$ of 
\eqref{16}, changes for $Z$ of \eqref{2} trivial. We first introduce a 
number of parameters. There is $\tilde{M}$ 
\begin{equation}\label{21}
	\tilde{M}=\frac{pN}4
\end{equation}
and $m_i$, for $i\geq 2$,
\begin{equation}\label{22}
	m_i=\frac{1}{i^{2+\epsilon i}}
\end{equation}
Each chunk is assigned a ``level'', a non-negative integer, and is either ``free''
or ``boxed''. It requires patience to develop the construction of these chunks.

We let $\mathcal{P}$ be the subset of indices for which
$\bar{J}_i\geq 0$ if $i\in\mathcal{P}$ and $\mathcal{N}$ be the subset
for which $\bar{J}_i<0$. Each chunk has a unique $\alpha_i$ assigned to the
$i\in\mathcal{P}$, say $\alpha_i=t_i$. Thus in a chunk some of the $\alpha_i$
may be summed over, but not the $\alpha_i$ with $i\in\mathcal{P}$. For
any chunk we define
\begin{equation}\label{23}
	R_0=\sum_{i\in\mathcal{P}}it_i
\end{equation}

\subsection*{Level-zero free chunks}
A level zero chunk is uniquely specified by the set of $t_i$, $i\in\mathcal P$. If
\begin{equation}\label{24}
	R_0\geq \tilde{M}
\end{equation}
then it is a free level-zero chunk. Its precise definition is
\begin{equation}\label{25}
	\prod_{i\in\mathcal P}(\subs(\alpha_i= t_i))\sum_{
	\stackrel{\alpha_i,i\in\mathcal N}{\sum_{i\in\mathcal N}i\alpha_i
	\leq\frac{pN}{2}-R_0}}\tilde{\beta}(N,\sum i\alpha_i)\left(
	\prod_i(\bar{J}_ip^iN)^{\alpha_i}\frac{1}{\prod(\alpha_i!)}\right)
\end{equation}
We are using a Mapple-like notation. In \eqref{25} the
$\alpha_i$ for $i\in\mathcal P$ are the set equal to $t_i$, and the
remaining $\alpha_i$ are summed over subject to the restriction from
\eqref{3}. As with all the chunks, this chunk is some subsum of the terms in 
\eqref{16}. Different chunks are disjoint, the union of 
all the chunks giving all terms in \eqref{16}.

\subsection*{Level-zero boxed chunks}
Here
\begin{equation}\label{26}
	R_0<\tilde{M}
\end{equation}
We define $C_0$ by
\begin{equation}\label{27}
	C_0=\sup_x\left\{x\mid\sum_{i\in\mathcal N}i\cdot
	\floor(xU_i)\leq\frac{pN}{2}-R_0\right\}
\end{equation}
where $\floor(\alpha)$ is the largest integer $\leq \alpha$. We
the set
\begin{equation}\label{28}
	m_0(i)=\floor(C_0U_i)
\end{equation}
The level-zero boxed chunk defined by the $t_i$, $i\in\mathcal P$ is then
\begin{equation}\label{29}
	\prod_{i\in\mathcal P}(\subs(\alpha_i=t_i))\sum_{\alpha_{a_1}=0}^{m_0(a_i)}
	\cdots\sum_{\alpha_{a_s}=0}^{m_0(a_s)}M
\end{equation}
Here $M$ indicates everything after the sums in \eqref{16}. $a_1,\dots,a_s$ are
the indices labelling elements of $\mathcal N$. By using \eqref{27} 
we have
found the biggest box, of a certain shape, we can insert in the
sum. That is we are picking the upper limits in \eqref{29} as large as possible,
with a certain fixed ratio between them

The $P_0\geq\tilde M$ in the free chunk will lead to 
enough smallness in estimates later that
one will not have to study cancellations between
signed terms via the contour integrals of \eqref{19}.
For the boxed chunks one will need to do so.

\subsection*{Level-one chunks}
The level-zero chunks fail to exhaust all terms
In \eqref{16} because of terms containing some $\alpha_i$ for some 
$i\in \mathcal N$ exceeding the upper limits in \eqref{29}.
We give a subset $\mathcal B_1$ of $\mathcal N$ and to each $i$
in $\mathcal B_1$ we associate a $t_i$ with
\begin{equation}\label{30}
t_i>m_0(i)
\end{equation}
We call the augmented set $\mathcal B_1$ of indices and associated $t_i$,
$\bar{\mathcal B}_1$. We set
\begin{equation}\label{31}
	R_1=\sum_{i\in\mathcal B_1}it_i
\end{equation}
If $R_0+R_1\geq\tilde{M}$ we have the level-one free chunk given as
\begin{equation}\label{32}
	\prod_{i\in\mathcal P\cup\mathcal B_1}(\subs(\alpha_i=t_i))
	\sum_{\stackrel{a_i,i\in\mathcal N_1}{\sum_{i\in\mathcal N_1}i\alpha_i
	\leq\frac{pN}{2}-R_0-R_1}}M
\end{equation}
We have defined $\mathcal N_1=\mathcal N-\mathcal B_1$. To define the boxed chunk
we define
\begin{equation}\label{33}
	C_1=\sup_x\left\{x\mid \sum_{i\in\mathcal N_1}i\floor(x U_i)\leq\frac{pN}{2}-
	R_0-R_1\right\}
\end{equation}
and 
\begin{equation}\label{34}
	m_1(i)=\floor(C_1 U_i)
\end{equation}
Then the level-one boxed chunk is given as 
\begin{equation}\label{35}
	\prod_{i\in\mathcal P\cup\mathcal B_1}(\subs(\alpha_i=t_i))
	\sum_{\alpha_{a_1}=0}^{m_1(a_1)}\cdots\sum_{\alpha_{a_s}=0}^{m_1(a_s)} M 
\end{equation}
and here the $a_1,\dots,a_s$ label elements of $\mathcal N_1$.

The set of $t_i$ associated to the $i$ in $\mathcal P\cap\mathcal B_1$
uniquely label the level-one chunks.

\subsection*{General level chunks}
We assume we have defined chunks of level-zero
through level-$n$, and we will derive expressions for the 
level-$(n+1)$ chunks. Thus we have
\begin{equation}\label{36}
	\mathcal N=\mathcal N_0\supset\mathcal N_1\supset
	\mathcal N_2\cdots\supset N_n
\end{equation}
\begin{equation}\label{37}
	\mathcal B_i\subset \mathcal N_{i-1},\qquad i=1,2,\dots,n
\end{equation}
\begin{equation}\label{38}
	\mathcal N_i=\mathcal N_i-\mathcal B_i,\qquad i=1,2,\dots,n
\end{equation}
\begin{equation}\label{39}
	\bar{\mathcal B}_i,\qquad i=1,2,\dots,n
\end{equation}
That is, we have an assignment $\alpha_i=t_i$ for 
$i$ in each $\mathcal B_i$. We set
\begin{equation}\label{40}
	R_k=\sum_{i\in\mathcal B_k}it_i,\qquad k=1,2,\dots,n
\end{equation}
and have $C_0,C_1,\dots, C_n$ with 
\begin{equation}\label{41}
	C_k=\sup_x\left\{x\mid \sum_{i\in\mathcal N_k}i\floor(xU_i)\leq
	\frac{pN}{2}-\sum_0^kRk\right\},\qquad k=0,1,\dots,n
\end{equation}
We set 
\begin{equation}\label{42}
	m_k(i)=\floor(C_k U_i),\qquad k=0,1,\dots,n
\end{equation}
and for $i$ in $\mathcal B_k$, $k=2,\dots,n$ one has
\begin{equation}\label{43}
	m_{k-1}(i)<t_i\leq m_{k-2}(i)
\end{equation}
For $k=1$ the analogous condition is given by \eqref{30};
there is in this case no upper bound here imposed. It is not difficult
to see the $C_i$ from a decreasing sequence.

To go to the next level, select the set $\mathcal B_{n+1}\subset \mathcal N_n$
and for $i\in\mathcal B_{n+1}$ require 
\begin{equation}\label{44}
	m_n(i)< t_0\leq m_{n-1}(i)
\end{equation}
The rest follows immediately.

Let us specify free and boxed chunks of level-$(n+1)$. The free level $n+1$ chunk
arises if
\begin{equation}\label{45}
	\sum_{i=0}^{n+1} R_i\geq \tilde{M}
\end{equation}
and then is given as
\begin{equation}\label{46}
	\prod_{i\in\mathcal P\cup(\mathcal N-\mathcal N_{n+1})}
	(\subs(\alpha_i=t_i))
	\sum_{\stackrel{\alpha_i,i\in\mathcal N_{n+1}}{\sum_{i\in\mathcal N_{n+1}}i\alpha_i
	\leq \frac{pN}{2}-\sum_{i=0}^{n+1}R_i}}M_i
\end{equation}
If \eqref{45} is not satisfied the boxed chunk is given as
\begin{equation}\label{47}
	\prod_{i\in\mathcal P\cup(\mathcal N-\mathcal N_{n+1})}(\subs(\alpha_i=t_i))
	\sum_{\alpha_{a_1}=0}^{m_{n+1}(a_1)}\cdots\sum_{\alpha_{a_s}=0}^{m_{n+1}(a_s)}M
\end{equation}
where $a_1,\dots,a_s$ are the indices of $\mathcal N_{n+1}$.

\section{Smallness from high occupation}\label{s6}
In this section we find an upper bound on 
\begin{equation}\label{48}
	\prod_i\left(\frac{(|\bar{J}_i|p^iN)^{\alpha_i}}{\alpha_i!}\right)
\end{equation}
subject to the restriction
\begin{equation}\label{49}
	\sum_ii\alpha_i\geq\tilde{M}
\end{equation}
We use \eqref{3}
\begin{equation}\label{50}
|\bar{J}_i|\leq r^i
\end{equation}
and define
\begin{equation}\label{51}
	Q=\prod_i\left(\frac{((pr)^iN)^{\alpha_i}}{\alpha_i!}\right)
\end{equation}
One then has
\begin{equation}\label{52}
	\ln Q\leq F= N\sum_i[ix_i\ln(pr)-x_i\ln x_i+x_i]
\end{equation}
using the well known inequality $n\ln(n/e)+1\leq \ln n!$,
where we have set
\begin{equation}\label{53}
	\alpha_i=x_iN
\end{equation}
\begin{equation}\label{54}
	\tilde{M}=\tilde{m} N
\end{equation}
Then we have
\begin{equation}\label{55}
	f=\frac{F}{N}=\sum_i[ix_i\ln(pr)-x_i\ln x_i+x_i]
\end{equation}
We ignore the restriction that $\alpha_i$ must be an integer. We note that 
our upper bound will apply if we restrict the set of
indices in the products of \eqref{48} or \eqref{51}, or
require the $\alpha_i$ to be integers, each of these would lead to a lower upper
bound. 

Now we want to maximize $f$ subject to 
\begin{equation}\label{56}
h=\sum_ix_i=\tilde{m}
\end{equation}
We apply Lagrange multipliers 
\begin{equation}\label{57}
	\frac{d}{dx_i}(f-\lambda h)=0
\end{equation}
getting
\begin{equation}\label{58}
	x_i=(pre^{-\lambda})^i
\end{equation}
and then, from \eqref{56}
\begin{equation}\label{59}
	\sum_{i=2}^\infty(pre^{-\lambda})^i=\tilde{m}
\end{equation}
one gets for $\tilde{m}=cp$ as $p$ goes to zero
\begin{equation}\label{60}
	F\cong-N\tilde{m}\vert\ln(p/\sqrt{\tilde{m}})\vert
\end{equation}

\section{The Contour Integrals, distorting the contours}\label{s7}
We have displayed the expressions for boxed  chunks of level-zero,
level-one, and general level-$(n+1)$
in eqs. \eqref{29}, \eqref{35}, \eqref{47}. We isolate the portion
of the expression shown in eq.~\eqref{18}, and relabelling indices look
at our present object or study
\begin{equation}\label{61}
	A=\sum_{\alpha_1=0}^{m_1}\cdots\sum_{\alpha_s=0}^{m_s}\prod_{i=1}^s
	\left(\frac{(J_{d_i}p^{d_i}N)^{\alpha_i}}{\alpha_i!}\right)\beta
\end{equation}
Here all the $J_i$ are negative. We set 
\begin{equation}\label{62}
	a_i=-J_{d_i}p^{d_i}N
\end{equation}
and rewrite $A$ as
\begin{equation}\label{63}
	A=\frac{1}{(2\pi i)^s}\oint_{C_1}dz_1\cdots\oint_{C_s}dz_s\prod_{i=1}^s\left(
	\frac{\pi}{\sin\pi z_i}\frac{a_i^{z_i}}{\Gamma(z_i+1)}\right)\beta
\end{equation}
where we take $C_i$ as hugging the interval 
$[-1/2,m_i+1/2]$, encircling it counterclockwise.

We let $g$ be the integrand of \eqref{63}
so that
\begin{equation}\label{64}
	A=\frac{1}{(2\pi i)^s}\oint_{C_1}dz_1\cdots\oint_{C_s}dz_s g
\end{equation}
and seek a stationary point of $g$, so we solve together
\begin{equation}\label{65}
	\frac{\partial}{\partial z_i} g=0,\qquad i=1,\dots,s
\end{equation}
We now use 
\begin{equation}\label{66}
	\Gamma(z)\Gamma(1-z)=\frac{\pi}{\sin\pi z}
\end{equation}
to write $g$ as
\begin{equation}\label{67}
	g=\prod_{i=1}^s\left(-a_i^{z_i}\Gamma(-z_i)\right)\beta
\end{equation}

We look at the equations \eqref{65} for a stationary point of $Z$,
where then $\beta=1$, getting
\begin{equation}\label{68}
	\frac{\partial}{\partial z_i}(-a_i^{z_i}\Gamma(-z_i))=0
\end{equation}
or taking exponentials and setting $w_i=-z_i$
\begin{equation}\label{69}
	\frac{d}{dw_i}[-w_i\ln a+\ln\Gamma(w)]=0
\end{equation}
We now use the standard large $w$ approximation
\begin{equation}\label{70}
	\ln n!\cong n\ln n-n
\end{equation}
to get
\begin{equation}\label{71}
	w_i\cong a_i+1
\end{equation}
or 
\begin{equation}\label{72}
	z_i=z_{i0}\cong-a_i-1
\end{equation}
at the stationary point.

In general let the coordinates of the stationary point be $z_i=z_{i0}$.
We then first shift the contour $C_i$ to $C_i'$ where $C_i'$
hugs the interval $[-1/2+\floor(z_{i0}),m_i+1/2]$
counterclockwise. One hits no singularities of 
$g$ to distort the contour so. Next we stretch the contour to infinity, plus and minus,
in the imaginary directions, leaving contours $C_i''$. 
$C_i''$ consists of two line segments parallel to the imaginary axis
\begin{equation}\label{73}
	\begin{gathered}
	\begin{aligned}
	C_i''&=[-\infty i-1/2+\floor(z_i),\infty i-1/2+\floor(z_i)]\\
	&\phantom{==}\mbox{union}\\
	&\phantom{==}[-\infty i+m_i+1/2,\infty i+m_i+1/2]
	\end{aligned}
	\end{gathered}
\end{equation}
Again it will not be hard to show that no singularities
of the integrand are crossed in the distortion, and the segments at $\infty$
to close the $C''_i$ contours contribute nothing.

In treating $Z^*$ and studying 
\begin{equation}\label{74}
	A=\frac{1}{(2\pi i)^s}\oint_{C''_i}dz_1\cdots\oint_{C''_s}dz_sg
\end{equation}
one will have to show the limit in eq~\eqref{17}
picks out just the value of the integrand at the 
center of the first line segment in \eqref{73}.
In studying $Z$ we will use a simpler route to
study \eqref{61}. (For one thing we do not want to 
worry about situations where $a_i$ is not large,
that is not guaranteed by \eqref{1}\dots\ perhaps there is no problem here at all.)

\section{Completion of the Proof for $Z$}\label{s8}
In this section we carry through the proof of the theorem, see eq.~\eqref{4},
for $Z$, much in the way we plan to complete the proof for $Z^*$, see
eq.~\eqref{17}. An important difference is that we do not need use of contour integration.
We divide $Z$ into a sum of three terms
\begin{equation}\label{75}
	Z=Z(N,p)=T_1+T_2+T_3
\end{equation}
$T_1(N,p)$ is the sum of level-zero boxed chunks, $T_2(N,p)$
is the sum of the rest of the boxed chunks, and $T_3(N,p)$ is 
the sum of the free chunks. We will arrive at the theorem by proving, for small
enough $p$, that
\begin{equation}\label{76}
	\lim_{N\to\infty}\frac{\ln T_1}{N}=\sum_2p^iJ_i
\end{equation}
\begin{equation}\label{77}
	\lim_{N\to\infty}\frac{\ln T_2}{N}<\sum_2p^iJ_i
\end{equation}
\begin{equation}\label{78}
	\lim_{N\to\infty}\frac{\ln T_3}{N}<\sum_2p^iJ_1
\end{equation}

\subsection*{Study of $T_3$}
We may overestimate $T_3$ as follows
\begin{equation}\label{79}
	T_3\leq\sum_{\alpha_i}Q
\end{equation}
where $Q$ is from \eqref{51} and the $\alpha_i$ are restricted by 
\begin{equation}\label{80}
	\sum i\alpha_i\geq \frac{pN}{4}
\end{equation}
Then we write
\begin{equation}\label{81}
	\begin{gathered}
	\begin{aligned}
	\sum_{\alpha_i}Q & =\sum_{\alpha_i}\prod_i\left(\frac{((pr)^iN)^{\alpha_i}}{
	\alpha_i!}\right)\\
	& \leq \sum_{\alpha_i}\prod_i\left(\frac{((pr)^iN)^{\alpha_i/2}}{(\alpha_i/2)!}\right)
	\prod_i\left(\frac{((pr)^iN)^{\alpha_i/2}}{(\alpha_i/2)!}\right)
	\end{aligned}
	\end{gathered}
\end{equation}
all subject to the restriction \eqref{80}. This yields
\begin{equation}\label{82}
	\sum_{\alpha_i}Q\leq AB
\end{equation}
with
\begin{equation}\label{83}
	A=\sup\prod_i\left(\frac{((pr)^iN)^{\alpha_i/2}}{(\alpha_i/2)!}\right)
\end{equation}
and
\begin{equation}\label{84}
B=\sum_i\prod_i\left(\frac{((pr)^iN)^{\alpha_i/2}}{(\alpha_i/2)!}\right)
\end{equation}
In \eqref{82} and \eqref{83} restriction \eqref{80} is enforced, but getting
a weaker inequality do not
impose it in \eqref{84}.
For $A$ from \eqref{60} we get 
\begin{equation}\label{85}
	A\cong e^{-N[WORDS]}
\end{equation}
Looking at $B$ we note
\begin{equation}\label{86}
\begin{aligned}
	\sum_i\frac{a^{i/2}}{(i/2)!} & =\sum_i\frac{a^i}{i!}+
	\sum_i\frac{a^{1/2+i}}{(1/2+i)!} \\
	& =\sum_i\frac{a^i}{i!}+
	a^{1/2}\sum_i\frac{a^{i}}{(1/2+i)!} \\
	& \leq(1+ca^{1/2})\sum \frac{a^i}{i!}\leq(1+ca^{1/2})e^a
\end{aligned}
\end{equation}
Equation \eqref{78} follows for $p$ small enough with little work.

\section{Dealing with $Z_1$ and $Z_2$}\label{s9}
In this section we study the methods used to treat $T_1$ and $T_2$ to derive
\eqref{76} and \eqref{77}. We discuss the proof of \eqref{76} alone,
since \eqref{77} requires no new ideas. We write $T_1$, the sum of the level zero
boxed chunks as
\begin{equation}\label{87}
	T_1=\sum_\beta A^\beta B^\beta
\end{equation}
with 
\begin{equation}\label{88}
	A^\beta=\prod_{i\in\mathcal P}\frac{(J_ip^iN)^{t_i}}{t_i!}
\end{equation}
where $t_i=t_i(\beta)$ satisfy
\begin{equation}\label{89}
	R_0=R_0(\beta)=\sum it_i<\frac{pN}{4}
\end{equation}
$C_0(\beta)$ is given by \eqref{27}, and $m_0(i)$ by \eqref{28}, using
which we define
\begin{equation}\label{90}
	B^\beta=\prod_{i\in\mathcal N}\left(\sum_{\alpha_i=0}^{m_0(i)}
	\frac{(J_ip^iN)^{\alpha_i}}{\alpha_i!}\right)
\end{equation}
We now set
\begin{equation}\label{91}
	B_0=\prod_{i\in\mathcal N}(e^{J_ip^iN})
\end{equation}
We write
\begin{equation}\label{92}
	A^\beta B^\beta=A^\beta B_0+A^\beta E^\beta
\end{equation}
and 
\begin{equation}\label{93}
	\sum A^\beta B^\beta=B_0\sum A^\beta+\sum A^\beta E^\beta
\end{equation}

We study the second term in the right side of \eqref{93} first
\begin{equation}\label{94}
	\left|\sum A^\beta E^\beta\right|\leq \left(\sum A^\beta\right)\sup_\beta
	|E^\beta|
\end{equation}
If we were studying $Z^*$ instead of $Z$ we would have used contour integral techniques,
but in the present case we use simpler methods
\begin{equation}\label{95}
	\sum A^\beta\leq e^{\sum_{i\in\mathcal P}J_ip^iN}
\end{equation}
To study $|E^\beta|$ we first introduce
\begin{equation}\label{96}
	g(a,n)=\sum_{i=n+1}^\infty\left(\frac{|a|^i}{i!}\right)e^{+|a|}
\end{equation}
and 
\begin{equation}\label{97}
	g_i=g(J_ip_i N,m_0(i))
\end{equation}
In terms of the quantities we have
\begin{equation}\label{98}
	|E^\beta|\leq e^{\sum_iJ_ip^iN}\cdot\left(\prod_{i\in\mathcal N}(1+g_i)-
	1\right)
\end{equation}
The dependence on $\beta$ comes from the dependence of $g_i$ on the $m_0(i)$
which depend on $\beta$.

If $x_i\geq 0$ and $\sum_i\leq 1$ one has teh inequality
\begin{equation}\label{99}
	\prod(1+x_i)-1\leq e\sum x_i
\end{equation}
One we know $\sum g_i\leq 1$ we now have
\begin{equation}\label{100}
	\left|\sum A^\beta E^\beta\right|\leq e^{\sum J_ip^iN}\cdot e
	\cdot \sup_\beta\left(\sum_{i\in\mathcal N}g_i\right)
\end{equation}

Limit \eqref{76} follows from 
\begin{equation}\label{101}
	\lim_{N\to\infty}\frac{1}{N}\ln\left(B_0\sum A^\beta\right)=\sum J_ip^i
\end{equation}
and
\begin{equation}\label{102}
	\lim_{N\to\infty}\frac{\left|\sum A^\beta E^\beta\right|}{B_0\sum A^\beta}=0
\end{equation}
We choose to see \eqref{101} by approximating $\sum A^\beta$ by its largest
term, a justifiable procedure in this case. Note that all terms
in sum are positive
This is the method used in \cite{1} to treat
\eqref{27}
there, and again in \cite{2} to treat~(5.24) therein.
In both these cases the method is used ``formally''
where it does not really apply since there are positive and negative
terms. It is the work of the present paper and its sequel, part~2, to
show the results in \cite{1} and \cite{2} are none the less correct.

We turn to understanding limit \eqref{102}.
From \eqref{100} we have
\begin{equation}\label{103}
	\frac{\left|\sum A^\beta E^\beta\right|}{B_0\sum A^\beta}\leq
	\frac{e^{\sum J_i p^i N}}{B_0\sum A^\beta}\cdot e\cdot
	\sup_\beta\left(\sum_{i\in\mathcal N}g_i\right)
\end{equation}
From \eqref{101}, now presumed true this becomes
\begin{equation}\label{104}
	\frac{\left|\sum A^\beta E^\beta\right|}{B_0\sum A^\beta}\leq
	e^{N\varepsilon(N)}\cdot e\cdot
	\sup_\beta\left(\sum_{i\in\mathcal N}g_i\right)
\end{equation}
for some $\varepsilon(N)$ that goes to zero with $N$. We may 
deduce an upper bound on $\sum_{i\in\mathcal N}g_i$ [WORDS]
$\sum_{i\in\mathcal N}\bar g_i$ with $\bar g_i$, $g_i$ computed with
$R_0=\frac{pN}{4}$.

We get an accurate upper bound for the $\sum_{i\in\mathcal N}\bar g_i$
by looking at
\begin{equation}\label{105}
	h(r^2p^2N,\gamma pN)
\end{equation}
with 
\begin{equation}\label{106}
	h(a,n)=\frac{a^n/n!}{e^{-a}}
\end{equation}
\begin{equation}\label{107}
	\ln h(r^2p^2N,\gamma pN)\leq -(\gamma pN)\ln
	\left(\frac{\gamma pN}{r^2p^2N}\right)+(\gamma pN)-r^2p^2N
\end{equation}
implying
\begin{equation}\label{108}
	h\leq e^{-\tilde\gamma N}
\end{equation}
with $\tilde\gamma>0$. This and \eqref{104} yield \eqref{102} and finally
\eqref{76}.

\end{document}